\documentclass[epj]{svjour}
\usepackage{graphics}
\usepackage{epstopdf}
\usepackage{amsmath}
\usepackage{ulem}
\usepackage{bm}
\begin{document}
\title{Dynamics of a polymer chain confined in a membrane}
\author{
S. Ramachandran\inst{1} 
\and 
S. Komura\inst{1}
\thanks{E-mail: komura@tmu.ac.jp}
\and 
K. Seki\inst{2}
\and 
G. Gompper\inst{3}
}                     % Do not remove

\institute{Department of Chemistry, 
Graduate School of Science and Engineering, 
Tokyo Metropolitan University, 
Tokyo 192-0397, Japan 
\and
National Institute of Advanced Industrial Science and Technology, 
Ibaraki 305-8565, Japan 
\and 
Institut f\"ur Festk\"orperforschung, 
Forschungszentrum J\"ulich, D-52425 J\"ulich, Germany, EU
}
\date{Received: date / Revised version: date}
\abstract{
We present a Brownian dynamics theory with full hydrodynamics 
(Stokesian dynamics)
for a Gaussian polymer chain embedded in a liquid membrane which is 
surrounded by bulk solvent and walls.
The mobility tensors are derived in Fourier space for the two geometries, 
namely, a free membrane embedded in a bulk fluid, and a membrane 
sandwiched by the two walls.
Within the preaveraging approximation, a new expression for the 
diffusion coefficient of the polymer is obtained for the free 
membrane geometry. 
We also carry out a Rouse normal mode analysis to obtain the 
relaxation time and the dynamical structure factor. 
For large polymer size, both quantities show Zimm-like behavior in 
the free membrane case, whereas they are Rouse-like for the sandwiched 
membrane geometry.
We use the scaling argument to discuss the effect of excluded volume 
interactions on the polymer relaxation time. 
 %end of PACS codes
} %end of abstract

\maketitle

%%%%%%%%%%%%%%%%%%%%%%%%%%%%%%%%%%%%%%%%%%%%%%%%%%%%
\section{Introduction}
\label{introduction}
%%%%%%%%%%%%%%%%%%%%%%%%%%%%%%%%%%%%%%%%%%%%%%%%%%%%

Integral membrane proteins play a vital role in a variety of 
cell functions such as solute transport, signal transduction
and regulation of membrane composition~\cite{alberts}.
Owing to finite temperatures, such proteins along with other 
membrane components are constantly undergoing Brownian motions.
The resulting diffusive motion plays an important role in 
determining their transport properties.
Hence the studies of diffusion constitutes an important basis 
for understanding the physical properties of membrane proteins,
and have been an active area of focus on model 
systems~\cite{peters-82,reitz-01,kahya-01,tsapis-01,gambin-07}
as well as on living 
cells~\cite{kucik-99,lippincott-01,vrljic-02,kenworthy-04}.

Although proteins consist of polymeric units of amino acids, the 
standard approach is to consider them as rigid disks moving 
in a two-dimensional (2D) liquid membrane under low-Reynolds 
number conditions.
The diffusion coefficient of a rigid disk translating in a membrane 
which is embedded in a three-dimensional (3D) bulk fluid 
was calculated by Saffman and Delbr\"uck 
(SD)~\cite{saffman-75,saffman-76}.
The obtained logarithmic size dependence is valid in the limit 
of small disk sizes. 
The SD theory was formally extended by Hughes {\it et al.} to all 
disk size ranges~\cite{hughes-81}.
In the case of large disk sizes, they showed that the diffusion 
coefficient is inversely proportional to its size, which is analogous 
to the Stoke-Einstein relation in 3D.
However, it should be noted that there is no single expression of the 
diffusion coefficient which covers the whole disk size ranges.
In a separate theoretical study, Evans and Sackmann (ES) employed a 
phenomenological approach to calculate the diffusion coefficient 
of a rigid disk moving in a membrane attached to a 
substrate~\cite{evans-88}.
The presence of a substrate in close proximity to the membrane was 
taken into account through a momentum decay term in the hydrodynamic 
equations.  
An extension of this work taking into account the effect of the advective
terms has also been done~\cite{Tserkovnyak-06}.
In addition to these, diffusion of rod shaped objects 
on membranes~\cite{levine-04,levine-04b} or on Langmuir 
monolayers~\cite{fischer-04} have also been theoretically analyzed.

Alternatively, the membrane protein can be regarded as a polymer 
chain rather than a rigid disk.
In this case, the internal degrees of freedom of the polymer should 
be taken into account, which is the main subject of this paper.
Apart from the protein analogy, hydrophobically modified polymers which 
adhere to the membrane could also be described using our 
description~\cite{yang-98}.
There are two theoretical works preceding the current work.
One of them is a study by Muthukumar on the dynamics of a hydrophobic 
polymer confined in a 2D liquid membrane~\cite{muthukumar-85}.
In his theory, the membrane itself was treated as an isolated 2D 
system having an anisotropic viscosity.
It was shown that the mean squared displacement of a monomer obeys
a diffusive law.
He also pointed out that the mode dependence of the relaxation time
arises from the excluded volume effect. 
The second and more direct precursor to the present work 
is the analytical calculation of the 
diffusion coefficient of a polymer chain~\cite{komura-95} using a 
2D hydrodynamic model with momentum 
decay~\cite{sriram-82,suzuki-89,seki-93,seki-07}.
Since their hydrodynamic model is essentially equivalent to that 
used by ES, the asymptotic size dependencies of the diffusion 
coefficient is the same between disks and polymers; namely, 
logarithmic in the small size limit, and algebraic in the 
large size limit.

Further motivation is provided by the experiments on DNA molecules 
embedded on a cationic supported membrane~\cite{maier-99,maier-00,maier-01}.
The negative charge of the DNA molecules leads to strong adhesion
with the membrane so that only the lateral motions are allowed.
The measured diffusion coefficient showed a Rouse-like behavior.
More recently, a similar experiment with DNA on a free standing 
membrane has been conducted~\cite{herold-10}.
Another related situation can be found in a dilute polymer solution 
confined between narrow slits.
Based on the scaling argument, such a polymer was predicted to show
a Rouse-like behavior~\cite{daoud-77,broch-77,degennes-broch-77,tlusty-06}.
In an attempt to verify these predictions, experiments on dilute 
solutions of DNA confined in narrow slits have shown that the 
exponents for conformation and chain relaxation of DNA 
is $2.2$ which lies between 2D and 3D behaviors~\cite{lin-07}.

\begin{figure}
\begin{center}
\resizebox{0.9\columnwidth}{!}{%
\includegraphics{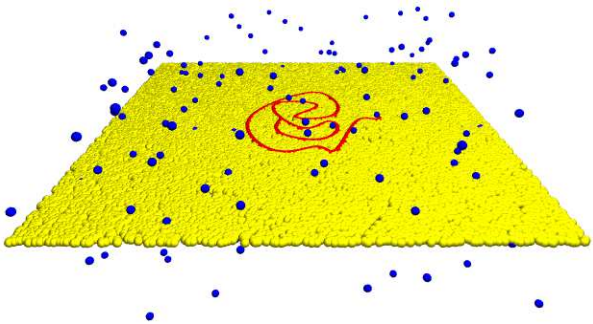}
}
\caption{A transmembrane protein approximated as a polymer chain (red chain)
embedded in a liquid membrane (yellow particles).
The membrane itself is surrounded by solvent (blue particles).
Only a few representative solvent particles are shown.}
\label{fig:concept}
\end{center}
\end{figure}

In this paper, we discuss a Brownian dynamics theory for a polymer 
chain confined in a membrane. 
As schematically presented in fig.~\ref{fig:concept}, we consider a
polymer chain (connected red particles) embedded in a liquid membrane 
(yellow particles) surrounded by a 3D bulk fluid (blue particles) and
walls (not shown).
We first derive the mobility tensors for the two geometries, i.e., a 
free membrane embedded in a bulk fluid, and a membrane sandwiched 
by two walls. 
For these two cases, we shall obtain the corresponding analytical
expressions for the polymer diffusion coefficient which are valid 
for all sizes.
We further perform a Rouse normal mode analysis to calculate the relaxation 
time and dynamical structure factor. 
For large polymer sizes, these quantities show Zimm-like behavior in 
the free membrane case, whereas they are Rouse-like in the presence 
of walls.
We also use a scaling theory to discuss the effect of excluded volume 
interactions on the relaxation times for the two geometries.
The present work demonstrates the importance of the outer environment 
of the membrane in determining the dynamics of a 2D polymer chain.

In the next section, we start to set up the governing equations for the 
membrane and the derivation of mobility tensors.
With the introduction of the polymer, the general formalism of the 
problem is constructed in sect.~\ref{polymer}. 
Sections~\ref{polySD} and~\ref{polyES} describe the results for the 
polymer dynamics for the free and confined membrane limiting cases, 
respectively.
Excluded volume effects are discussed using scaling arguments in
sect.~\ref{excV}.
We finally close with several discussions in sect.~\ref{discussion}.

%%%%%%%%%%%%%%%%%%%%%%%%%%%%%%%%%%%%%%%%%%%%%%%%%%%%
\section{Membrane hydrodynamics}
\label{membrane}
%%%%%%%%%%%%%%%%%%%%%%%%%%%%%%%%%%%%%%%%%%%%%%%%%%%%

Before introducing the polymer, we first establish 
the governing equations for the membrane and its surrounding environment.
The aim of this section is to derive the membrane mobility tensors
which will be used in the later sections for the polymer equations of 
motion.
The present calculation closely follows the formulation by Inaura and 
Fujitani~\cite{inaura-08}.
However we consider a more general situation which will be described below.
The details of the calculation are relegated to appendix A.

\begin{figure}%[h!t]
\begin{center}
\resizebox{0.9\columnwidth}{!}{
\includegraphics{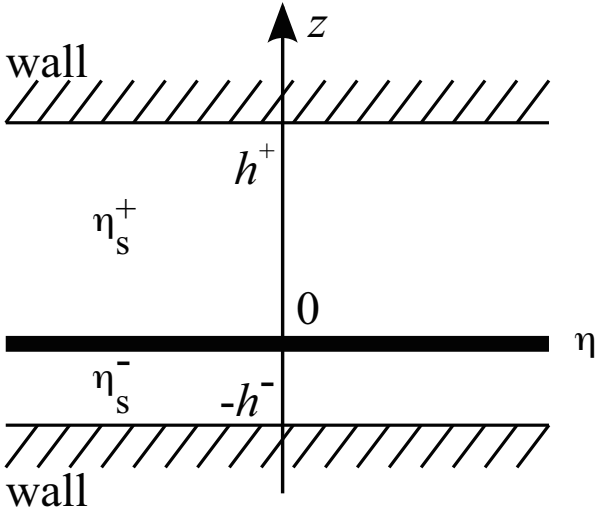}
}
\caption{Schematic picture showing a planar liquid membrane having 
2D viscosity $\eta$ located at $z=0$.
It is sandwiched by a solvent of 3D viscosity $\eta_{\rm s}^{\pm}$.
Two impenetrable walls are located at $z=\pm h^\pm$ bounding the 
solvent.} 
\end{center}
\label{fig:membwall}
\end{figure}

As shown in fig.~\ref{fig:membwall}, we assume that the membrane is 
an infinite planar sheet of liquid, and its out-of-plane fluctuations 
are totally neglected, which is justified for typical bending rigidities
of bilayers.
Relaxation dynamics of membrane fluctuations near 
walls have been previously considered~\cite{kraus-94,gov-04,sumithra-02}.
The liquid membrane is embedded in a bulk fluid such as water or 
solvent which is bounded by hard walls.
Let ${\bf v}({\bf r})$ be the 2D velocity of the membrane fluid 
and the 2D vector ${\bf r}=(x,y)$ represents a point in the plane 
of the membrane.
We first assume the membrane to be incompressible  
\begin{equation}
\nabla \cdot {\bf v} = 0, 
\label{eqn:2Dcompress}
\end{equation}
where $\nabla$ is a 2D differential operator.
We work in the low-Reynolds number regime of the membrane hydrodynamics
so that the inertial effects can be neglected.
This allows us to use the 2D Stokes equation given by 
\begin{equation}
\eta \nabla^2 {\bf v} - \nabla p + {\bf f}_{\rm s} + \bf{F} =0,
\label{eqn:2Dstokes}
\end{equation}
where $\eta$ is the 2D membrane viscosity,
$p({\bf r})$ the 2D in-plane pressure,
${\bf f}_{\rm s}({\bf r})$ the force exerted on the membrane by the 
surrounding fluid (``s'' stands for the solvent), 
and ${\bf F}({\bf r})$ is any other force acting on the membrane such 
as that due to a polymer chain introduced in sect.~\ref{polymer}.

As presented in fig.~\ref{fig:membwall}, the membrane is fixed in the 
$xy$-plane at $z=0$.
The upper ($z>0$) and the lower ($z<0$) fluid regions are denoted
by $``+"$ and $``-"$, respectively. 
The velocities and pressures in these regions are written as 
${\bf v}^{\pm}({\bf r},z)$ and $p^{\pm}({\bf r},z)$, respectively.
Since the 3D viscosity of the upper and the lower solvent can be 
different, we denote them as $\eta_{\rm s}^{\pm}$ respectively. 
Consider the situation in which impenetrable walls are located 
at $z=\pm h^{\pm}$, where $h^+$ and $h^-$ can be different in general.
We note that this point differs from ref.~\cite{inaura-08}.
Similar to the liquid membrane, the solvent in both regions
are taken to be incompressible 
\begin{equation}
\tilde{\nabla} \cdot {\bf v}^{\pm}  = 0,
\label{eqn:3Dcompress}
\end{equation}
where $\tilde \nabla$ represents a 3D differential operator.
We also neglect the solvent inertia and hence the solvent obeys 
the 3D Stokes equations 
\begin{equation}
\eta_{\rm s}^{\pm} \tilde{\nabla}^2 {\bf v}^{\pm} - 
\tilde{\nabla} p^{\pm} = 0.
\label{eqn:3Dstokes}
\end{equation}
The presence of the surrounding solvent is important because it 
exerts force on the liquid membrane.
This force, indicated as ${\bf f}_{\rm s}$ in eq.~(\ref{eqn:2Dstokes}), 
is given by the projection of 
$({\bm \sigma}^+ - {\bm \sigma}^-)_{z=0} \cdot \hat{{\bf e}}_z$
on the $xy$-plane.
Here $\hat{{\bf e}}_z$ is the unit vector along the $z$-axis, 
and ${\bm \sigma}^{\pm}$ are the stress tensors due to the solvent
\begin{equation}
{\bm \sigma}^{\pm} = 
- p^{\pm} {\bf I} 
+ \eta_{\rm s}^{\pm} [\tilde{\nabla} {\bf v}^{\pm} 
+ (\tilde{\nabla} {\bf v}^{\pm})^{\rm T}].
\label{eqn:stress}
\end{equation}
In the above, ${\bf I}$ is the identity tensor and the superscript
``T'' indicates the transpose.

Using the stick boundary conditions at $z=0$ and $z=\pm h^\pm$, we 
solve the hydrodynamic equations (\ref{eqn:3Dcompress}) and 
(\ref{eqn:3Dstokes}) to obtain ${\bf f}_{\rm s}$. 
Then we calculate the membrane velocity from 
eq.~(\ref{eqn:2Dstokes}) as  
\begin{equation}
{\bf v}[{\bf k}]
= {{\bf G}}[{\bf k}] \cdot {\bf F}[{\bf k}],
\label{eqn:membvelFT}
\end{equation}
where ${\bf v}[{\bf k}]$ and ${\bf F}[{\bf k}]$ are the Fourier 
components of ${\bf v}({\bf r})$ and ${\bf F}({\bf r})$ defined by
\begin{equation}
{\bf v}({\bf r}) = 
\int \frac{{\rm d}^2 k}{(2\pi)^2}
{\bf v}[{\bf k}]
\exp ( i {\bf k}\cdot {\bf r}),
\label{eqn:FT}
\end{equation}
and
\begin{equation}
{\bf F}({\bf r}) = 
\int \frac{{\rm d}^2 k}{(2\pi)^2}
{\bf F}[{\bf k}]
\exp ( i {\bf k}\cdot {\bf r}),
\end{equation} 
respectively, and ${\bf k}=(k_x,k_y)$.
After some calculations (see appendix A for the details), one
can show that the mobility tensor ${\bf G}[{\bf k}]$ in Fourier 
space is given by 
\begin{align}
G_{\alpha\beta}[{\bf k}] = &  
\frac{1}{\eta k^2 + k[ \eta_{\rm s}^+ \coth(kh^+)
+\eta_{\rm s}^-\coth(kh^-)]} 
\nonumber\\
& \times \left( \delta_{\alpha\beta} - \frac{k_\alpha k_\beta}{k^2} \right),
\label{eqn:genoseen}
\end{align}
with $\alpha, \beta = x,y$ and $k = \vert {\bf k} \vert$.
As in ref.~\cite{inaura-08}, we mainly consider (except in 
sect.~\ref{discussion}) the case when the two walls are located 
at equal distances from the membrane, i.e., $h^+=h^-=h$.
Then the above mobility tensor becomes
\begin{equation}
G_{\alpha\beta}[{\bf k}] =  \frac{1}{\eta [ k^2 + \nu k \coth(kh)]} 
\left( \delta_{\alpha\beta} - \frac{k_\alpha k_\beta}{k^2} \right),
\label{eqn:cothoseen}
\end{equation}
where $\nu \equiv 2 \eta_{\rm s} / \eta$ with 
$\eta_{\rm s} = (\eta_{\rm s}^{+}+\eta_{\rm s}^{-})/2$.
An almost equivalent expression to eq.~(\ref{eqn:cothoseen}) has also 
been derived for Langmuir monolayers in which there is only one 
wall or a substrate~\cite{fischer-04,lubensky-96}.
In the following, we discuss the two limiting situations of 
eq.~(\ref{eqn:cothoseen}).

Saffman and Delbr\"uck (SD) investigated the case when the two walls 
are located infinitely away from the membrane~\cite{saffman-75,saffman-76}. 
This is called as the free membrane case and all the 
related physical quantities are denoted by the superscript ``SD''.  
Taking the limit of $kh\gg1$ in eq.~(\ref{eqn:cothoseen}), the 
mobility tensor becomes~\cite{inaura-08,oppenheimer-09,oppenheimer-10}
\begin{equation}
G^{\rm SD}_{\alpha\beta}[{\bf k}] =  
\frac{1}{\eta ( k^2 + \nu k)} 
\left( \delta_{\alpha\beta} - \frac{k_\alpha k_\beta}{k^2} \right).
\label{eqn:sdoseen}
\end{equation}
Notice that the length $\nu^{-1}$ is called the SD hydrodynamic 
screening length.
The real space expression of this mobility tensor is obtained by the 
Fourier transform of eq.~(\ref{eqn:sdoseen})
\begin{equation}
{\bf G}({\bf r}) = \int \frac{{\rm d}^2k}{(2 \pi)^2}
{\bf G}[{\bf k}] \exp (i {\bf k}\cdot{\bf r}).
\label{eqn:Greal}
\end{equation}
Performing the calculations presented in appendix B, we 
obtain~\cite{fischer-04,oppenheimer-09,oppenheimer-10}
\begin{align}
G^{\rm SD}_{\alpha\beta}({\bf r}) 
= & \frac{1}{4\eta} \left[ {\bf H}_0(\nu r) - Y_0(\nu r)
+\frac{2}{\pi\nu^2r^2} \right. 
\nonumber\\
& \left.-
\frac{{\bf H}_1(\nu r)}{\nu r} +\frac{Y_1(\nu r)}{\nu r}\right]
\delta_{\alpha\beta} 
\nonumber\\
+ & \frac{1}{4\eta} \left[ -\frac{4}{\pi\nu^2r^2}
+\frac{2{\bf H}_1(\nu r)}{\nu r}\right.
\nonumber\\ 
&\left.-
\frac{2Y_1(\nu r)}{\nu r} -{\bf H}_0(\nu r) + Y_0(\nu r) \right]
\frac{r_\alpha r_\beta}{r^2},
\label{eqn:realGsd}
\end{align}
where $r = \vert{\bf r}\vert$.
In the above, ${\bf H}_n(z)$ are the Struve functions and 
$Y_n(z)$ are the Neumann functions or the Bessel functions of 
the second kind.

In the opposite $kh\ll 1$ limit, the membrane is confined between
the two walls.
Since such a limiting case was considered by Evans and Sackmann 
(ES)~\cite{evans-88}, we denote all the physical quantities for 
this situation with the superscript ``ES''.
In this case, eq.~(\ref{eqn:cothoseen}) takes the following form
\begin{equation}
G^{\rm ES}_{\alpha\beta}[{\bf k}] =  
\frac{1}{\eta ( k^2 + \kappa^2)} 
\left( \delta_{\alpha\beta} - \frac{k_\alpha k_\beta}{k^2} \right), 
\label{eqn:esoseen}
\end{equation}
where $\kappa \equiv(\nu/h)^{1/2}$.
This new length scale $\kappa^{-1}$ is the ES hydrodynamic screening 
length.
We note that $\kappa^{-1}$ is the geometric mean of $\nu^{-1}$ and 
$h$~\cite{stone-98}.   
The above ES mobility tensor was previously used in a 
phenomenological membrane hydrodynamic 
model~\cite{komura-95,seki-93,seki-07,oppenheimer-10,sanoop-drag-10}.
Following the calculations in appendix B, the real space representation 
of the ES mobility tensor becomes 
\begin{align}
G^{\rm ES}_{\alpha\beta}({\bf r})&=
\frac{1}{2\pi\eta} \left[ K_0(\kappa r) 
+ \frac{K_1(\kappa r)}{\kappa r} 
- \frac{1}{\kappa^2 r^2} \right]
\delta_{\alpha\beta}
\nonumber\\
&+ \frac{1}{2\pi\eta} \left[ -K_0(\kappa r) 
- \frac{2K_1(\kappa r)}{\kappa r} 
+ \frac{2}{\kappa^2 r^2} \right]
\frac{r_\alpha r_\beta}{r^2},
\label{eqn:realGes}
\end{align}
where $K_n(z)$ are the modified Bessel functions of the second kind.
We use either eq.~(\ref{eqn:sdoseen}) or eq.~(\ref{eqn:esoseen})
in our subsequent calculations with a polymer.

Strictly speaking, eqs.~(\ref{eqn:realGsd}) and~(\ref{eqn:realGes})
should have been obtained by taking the limits of $\nu h \gg 1$ and 
$\nu h \ll 1$, respectively, after the integration of 
eq.~(\ref{eqn:cothoseen}) over ${\bf k}$. 
However the inverse Fourier transform of eq.~(\ref{eqn:cothoseen}) is 
nontrivial. 
The present approach serves our purpose and the rigorous derivation
will be given in a separate publication.

%%%%%%%%%%%%%%%%%%%%%%%%%%%%%%%%%%%%%%%%%%%%%%%%%%%%%%%%%%%%%%
\section{Dynamics of a 2D Gaussian polymer chain embedded in 
a membrane}
\label{polymer}
%%%%%%%%%%%%%%%%%%%%%%%%%%%%%%%%%%%%%%%%%%%%%%%%%%%%%%%%%%%%%%

We are now ready to introduce a polymer into the membrane.
For simplicity, we first work with a Gaussian polymer chain whose 
conformation is given by a set of $N$ position vectors denoted as 
$\{ {\bf R}_n \} = ({\bf R}_1, \ldots, {\bf R}_N)$ embedded 
in the 2D membrane. 
The excluded volume effects will be discussed later in sect.~\ref{excV}.
It is implicitly assumed that the polymer relaxation time is much 
longer than that for the typical hydrodynamic disturbances so that 
the membrane can be effectively considered as a 2D liquid.
If the polymer consists of monomers which exert a set of point 
forces ${\bf f}_n$ acting at ${\bf R}_n$, the external force 
due to the polymer ${\bf F}({\bf r})$ in eq.~(\ref{eqn:2Dstokes}) 
can be written as
\begin{equation}
{\bf F}({\bf r}) = \sum_n {\bf f}_n \, \delta({\bf r} - {\bf R}_n).
\label{eqn:Fpolytot}
\end{equation}
In writing this expression, we have assumed that the superposition 
principle holds. 
With the use of the mobility tensor obtained in the previous section, 
eq.~(\ref{eqn:2Dstokes}) can be formally solved as  
\begin{equation}
{\bf v}({\bf r}) = \sum_n {\bf G}({\bf r} - {\bf R}_n) 
\cdot {\bf f}_n.
\end{equation}
Since monomers move with the same velocity as the membrane, their
velocities are given by 
\begin{equation}
\frac{\partial{\bf R}_n(t)}{ \partial t} = {\bf v}({\bf R}_n) 
= \sum_m {\bf G}_{nm} \cdot {\bf f}_m,
\end{equation}
where we have used the notation
${\bf G}_{nm}\equiv{\bf G}({\bf R}_n - {\bf R}_m)$.

The Langevin equation for a polymer chain embedded in a membrane is 
written as~\cite{ermak-78,doi-edwards}
\begin{align}
\frac{\partial {\bf R}_n(t)}{\partial t}
&= \sum_{m} {\bf G}_{nm} \cdot 
\left(
-\frac{\partial U}{\partial {\bf R}_m}
+ {\bm\zeta}_m(t)
\right)
\nonumber\\
&+\frac{k_{\rm B} T}{2} \sum_m 
\frac{\partial }{\partial {\bf R}_m} 
\cdot {\bf G}_{nm},
\label{eqn:langevin}
\end{align}
where ${\bm \zeta}_m(t)$ is the Gaussian random force acting at 
${\bf R}_m$, $k_{\rm B}$ the Boltzmann constant, $T$ 
the temperature.  
The potential energy of the 2D polymer has the form
\begin{equation}
U= \frac{ k_{\rm B}T}{b^2} 
\sum_{n=2}^N ({\bf R}_n - {\bf R}_{n-1})^2,
\label{eqn:polypot}
\end{equation}
where $b$ is the Kuhn length.
It can be shown that the mobility tensors eqs.~(\ref{eqn:realGsd}) 
and (\ref{eqn:realGes}) satisfy 
\begin{equation}
\frac{\partial}{\partial {\bf R}_m} \cdot {\bf G}_{nm} = 0.
\label{eqn:divG}
\end{equation}
Substituting eq.~(\ref{eqn:polypot}) into eq.~(\ref{eqn:langevin}) 
and using eq.~(\ref{eqn:divG}), we have
\begin{equation}
\frac{\partial {\bf R}_n(t)}{\partial t}  =
\sum_{m} {\bf G}_{nm}
\cdot \left(
\frac{ 2k_{\rm B} T}{b^2} 
\frac{\partial^2 {\bf R}_m(t)}{\partial m^2}
+{\bm\zeta}_m(t)
\right).
\label{eqn:monomereq}
\end{equation}

Due to the hydrodynamic coupling between different parts of the 
polymer, the above equation is non-linear and difficult to solve 
analytically. 
In order to overcome this difficulty, we employ the preaveraging 
approximation which has been successfully used for a polymer in 3D 
solvent~\cite{doi-edwards,zimm-56}. 
Assuming that the polymer is close to its equilibrium, we replace 
${\bf G}_{nm}$ by its equilibrium value 
$\langle {\bf G}_{nm} \rangle$ such that
\begin{align}
&\langle {\bf G}_{nm}\rangle =
\int {\rm d}\{{\bf R}_n\} \, {\bf G}_{nm} 
{\rm \Psi}(\{{\bf R}_n\}) 
\nonumber \\
&=\int_0^\infty {\rm d}r \,
2\pi r
 \frac{1} {\pi \vert n-m \vert b^2}
\exp \left( - \frac{r^2}{\vert n-m \vert b^2} \right)
{\bf G}_{nm}(r)
\nonumber \\
& = g(n-m){\bf I},
\end{align}
where ${\rm \Psi}(\{{\bf R}_n\})$ is the 2D Gaussian distribution 
function.
Within this approximation, eq.~(\ref{eqn:monomereq}) can be simplified 
as    
\begin{equation}
\frac{\partial {\bf R}_n(t)}{\partial t} 
 = \sum_m g(n-m)
\left(
\frac{ 2 k_{\rm B} T}{b^2}
\frac{\partial^2 {\bf R}_m(t)}{\partial m^2} 
+ {\bm\zeta}_m(t)
\right).
\label{eqn:polyeom}
\end{equation}

The above equation can be rewritten in terms of the Rouse normal 
coordinates defined by~\cite{doi-edwards} 
\begin{equation}
{\bf X}_p(t) = \frac{1}{N}\int_0^N {\rm d}n 
\cos\left( \frac{p \pi n}{N}\right)
{\bf R}_n(t),
\label{eqn:Rouse}
\end{equation}
as 
\begin{equation}
\frac{\partial {\bf X}_p(t)}{\partial t} 
= \sum_q g_{pq} [ - k_q {\bf X}_q (t)+ {\bm\zeta}_q(t)],
\end{equation}
with 
\begin{equation}
k_p = \frac{4 \pi^2 k_{\rm B}T}{Nb^2}p^2.
\end{equation}
Here $p, q=0,1,2,\ldots$ and $g_{pq}$ is the mobility tensor in 
terms of the normal coordinates: 
\begin{equation}
g_{pq} = \int_0^N \frac{{\rm d}n}{N}  
\int_0^N \frac{{\rm d}m}{N} 
\cos \left( \frac{p \pi n}{N} \right)
\cos \left( \frac{q \pi m}{N} \right) g(n-m).
\label{eqn:gengpq}
\end{equation}
If one neglects the contribution from the off-diagonal components 
of $g_{pq}$, we finally obtain 
\begin{equation}
\frac{\partial {\bf X}_p(t)}{\partial t} = 
g_{p}\left[-k_p {\bf X}_p(t) + {\bm\zeta}_p(t) \right],
\end{equation}
with the definition $g_p \equiv g_{pp}$.
The Gaussian random forces ${\bm\zeta}_p(t)$ satisfy the following 
conditions  
\begin{align}
& \langle \zeta_{p\alpha}(t)\rangle =  0, \\
& \langle \zeta_{p\alpha}(t)\zeta_{q\beta}(t') \rangle 
= 2 \delta_{pq}\delta_{\alpha\beta} (g_{p})^{-1} 
k_{\rm B} T \delta(t-t').
\end{align}
Therefore the relaxation time of a polymer in terms of the Rouse
modes is given by 
\begin{equation}
\tau_p=\frac{1}{g_{p}k_p}.
\label{eq:taup}
\end{equation}
Furthermore, the polymer diffusion coefficient can be  
calculated according to the following equation
\begin{equation}
D = k_{\rm B}T g_{0}= 
k_{\rm B}T \int_0^N  \frac{{\rm d}n}{N} \int_0^N  \frac{{\rm d}m}{N} 
\, g(n-m).
\label{eqn:genD}
\end{equation}

Another useful quantity that we calculate is the dynamic structure 
factor defined by
\begin{equation}
S({\bf k},t)=\frac{1}{N} 
\sum_{n,m} 
\langle
\exp [ i {\bf k} \cdot ( {\bf R}_n(t) - {\bf R}_m(0) )]
\rangle.
\label{eq:skt}
\end{equation}
Since ${\bf R}_n(t) - {\bf R}_m(0)$ is a linear function of 
${\bm\zeta}_n(t)$ obeying the Gaussian distribution, the 
distribution of ${\bf R}_n(t) - {\bf R}_m(0)$ is also 
Gaussian~\cite{feller}.
Hence we have 
\begin{align}
& \langle \exp(i {\bf k}\cdot [{\bf R}_n(t) - {\bf R}_m(0)])\rangle
\nonumber \\ 
& = \exp\left(- \frac{k^2}{4} \langle ( {\bf R}_n(t) - 
{\bf R}_m(0) )^2 \rangle  \right),
\end{align}
in 2D.
Denoting the center of mass by ${\bf X}_0$ and using the inverse 
relation of eq.~(\ref{eqn:Rouse}) 
\begin{equation}
{\bf R}_n = {\bf X}_0 
+ 2 \sum_{p=1}^\infty
{\bf X}_p \cos\left(
\frac{p \pi n}{N} \right),
\end{equation}
we can calculate the dynamical structure factor as 
\begin{align}
& S({\bf k},t) = \frac{1}{N} \sum_{n,m}
\exp \Big[
-k^2 D t
- \frac{1}{4}\vert n - m \vert b^2 k^2
\nonumber\\
& -\frac{Nb^2k^2}{\pi^2}
\sum_{p=1}^\infty
\frac{1}{p^2}
\cos\left( \frac{p \pi n}{N}\right)
\cos\left( \frac{p \pi m}{N}\right)
[1 - \exp(-t/\tau_p)]
\Big].
\end{align}
This completes the general formalism of the 2D polymer dynamics 
confined in a liquid membrane.
In the next section, we consider the free and confined membrane 
cases separately by using eqs.~(\ref{eqn:sdoseen}) or (\ref{eqn:esoseen}) 
for the mobility tensor.

%%%%%%%%%%%%%%%%%%%%%%%%%%%%%%%%%%%%%%%%%%%%%%%%%%%%
\section{Polymer dynamics: free membrane case}
\label{polySD}
%%%%%%%%%%%%%%%%%%%%%%%%%%%%%%%%%%%%%%%%%%%%%%%%%%%%

When the two walls in fig.~\ref{fig:membwall} are located 
at infinite distance from the membrane, one can use the SD 
mobility tensor given by eq.~(\ref{eqn:sdoseen}). 
We calculate the preaveraged mobility tensor, relaxation time, 
diffusion coefficient and structure factor following the recipe 
described in the previous section.

%%%%%%%%%%%%%%%%%%%%%%%%%%%%%%%%%%%%%%%%%%%%%%%%%%%%
\subsection{Mobility tensor}

In order to perform the preaveraging of the mobility tensor, 
we take the configurational average of eq.~(\ref{eqn:sdoseen}) 
by using the equilibrium probability distribution function of a 
Gaussian polymer in 2D:
\begin{align}
\langle {\bf G}^{\rm SD}_{nm}\rangle
&= \left\langle
\int \frac{ {\rm d}^2 k}{(2 \pi)^2}
\frac{ {\mathbf I} -  \hat {\bf k} \hat {\bf k}}
{\eta(k^2 + \nu k)}
\exp [i {\bf k} \cdot ({\bf R}_n - {\bf R}_m )]
\right\rangle, 
\end{align}
where $\hat {\bf k}$ denotes a unit vector along ${\bf k}$. 
This leads to $\langle {\bf G}_{nm}^{\rm SD} \rangle
= g^{\rm SD}(n-m) {\bf I}$ with
\begin{align}
&g^{\rm SD}(n-m) \nonumber \\
&= \frac{1}{4 \pi \eta} \int_0^\infty {\rm d}k \, 
\frac{k}{ k^2 + \nu k}
\exp\left(-\frac{1}{4} b^2 k^2 \vert n-m \vert \right)
\nonumber\\
&=\frac{1}{8\pi\eta}
\exp\left( - \frac{1}{4}b^2 \nu^2 \vert n-m \vert \right)
\nonumber\\
&\times \left[
\pi{\rm erfi} \left( \frac{1}{2}b\nu \sqrt{\vert n-m \vert} \right)
- {\rm Ei}\left( \frac{1}{4} b^2 \nu^2 \vert n-m \vert \right)
\right],
\label{eqn:sdmobility}
\end{align}
where ${\rm erfi}(z)$ is the imaginary error function 
\begin{equation}
{\rm erfi}(z) = -i{\rm erf}(iz),
\end{equation}
with ${\rm erf}(z)$ being the error function
\begin{equation}
{\rm erf}(z) = \frac{2}{\sqrt{\pi}}\int_0^z {\rm d}u \, e^{-u^2},
\end{equation}
whereas ${\rm Ei}(-z)$ is the exponential integral
given by~\cite{abram-stegun}  
\begin{equation}
{\rm Ei}(-z) = -\int_z^\infty {\rm d} u \, \frac{e^{-u}}{u}.
\end{equation}
It should be noted that the obtained $g^{\rm SD}(n-m)$ is real 
despite the presence of complex functions.

%%%%%%%%%%%%%%%%%%%%%%%%%%%%%%%%%%%%%%%%%%%%%%%%%%%%
\subsection{Relaxation time}

In order to obtain the polymer relaxation time, we first 
substitute eq.~(\ref{eqn:sdmobility}) into eq.~(\ref{eqn:gengpq}) 
to express the mobility tensor in terms of the Rouse normal
coordinates
\begin{align}
&g^{\rm SD}_{p}=
\frac{1}{\pi \eta Nb^2}
\int_0^\infty {\rm d}k \,
\frac{k^3}
{(k^2 + \nu k)
[ k^4 + \left( 4 \pi p/N b^2 \right)^2 ]}
\nonumber\\
&=\frac{1}{16 \pi \eta }
\frac{ 
\pi^2 p
-  \sqrt{2p} \pi^{3/2}\delta
+ 2 \ln(\pi p/\delta^2)\delta^2
+( \sqrt{2 \pi/p})\delta^3 
}
{
\pi^2 p^2
+\delta^4  
},
\end{align}
(note that $g^{\rm SD}_{p}=g^{\rm SD}_{pp}$).
In the above, we have defined the dimensionless polymer size 
$\delta \equiv \sqrt{N}b \nu/2$.
Since the radius of gyration for the 2D Gaussian polymer is 
$R_{\rm g} = \sqrt{N}b/2$~\cite{doi-edwards}, 
$\delta$ can be also written as 
$\delta = R_{\rm g} \nu$. 
Using eq.~(\ref{eq:taup}), the relaxation time becomes  
\begin{align}
& \tau_p^{\rm SD}= \frac{4 N b^2 \eta}{\pi k_{\rm B}T}
\nonumber \\
& \times
\frac{
\pi^2 p^2+ 
\delta^4 
}
{p^2[
\pi^2 p
-  \sqrt{2p}\pi^{3/2} \delta
+2 \ln(\pi p/\delta^2)\delta^2
+(\sqrt{2 \pi/p}) \delta^3 
]}.
\label{eqn:tausd}
\end{align}
This expression shows how the presence of the bulk solvent 
affects the relaxation time.

We consider two asymptotic limits of eq.~(\ref{eqn:tausd}).
For small polymer sizes or $\delta \ll 1$, we have
\begin{equation}
\tau_p^{\rm SD}
\approx \frac{4 N b^2 \eta}{\pi k_{\rm B}T} \frac{1}{p}.
\label{eqn:tausdsmalld}
\end{equation}
For large sizes, the condition $\delta \gg 1$ yields
\begin{equation}
\tau_p^{\rm SD}
\approx \frac{4 N b^2 \eta}{\pi k_{\rm B}T} 
\frac{\delta}{\sqrt{2\pi}p^{3/2}}.
%= \frac{4\sqrt{2} N b^2 \eta_{\rm s} R_{\rm g}}{\pi^{3/2} k_{\rm B}T p^{3/2}}.
\label{eqn:tausdlarged}
\end{equation}
It should be noticed that eq.~(\ref{eqn:tausdlarged}) 
depends only on the solvent viscosity $\eta_{\rm s}$ but not on 
the membrane viscosity $\eta$.

\begin{figure}
\begin{center}
\resizebox{0.9\columnwidth}{!}{%
\includegraphics{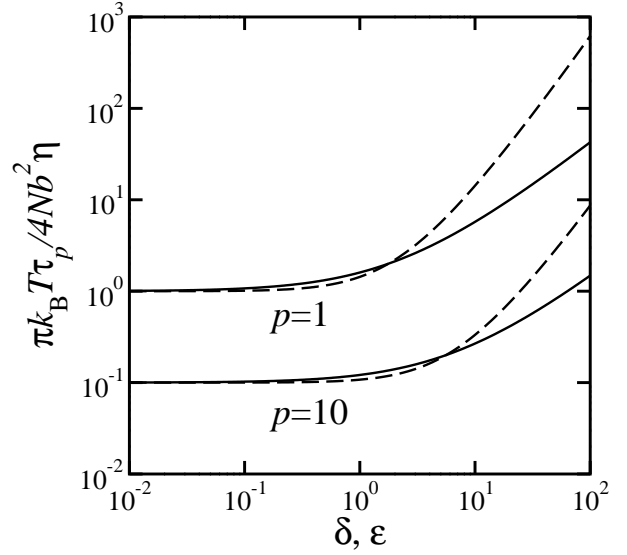}
}
\caption{Scaled relaxation time $\pi k_{\rm B}T \tau_p /4 N b^2 \eta$
as a function of $\delta = R_{\rm g} \nu$ for free membranes (solid lines) 
or $\varepsilon = R_{\rm g} \kappa$ for confined membranes (dashed lines)
for $p=1$ and $10$.}
\label{fig:taup}
\end{center}
\end{figure}
\begin{figure}
\begin{center}
\resizebox{0.9\columnwidth}{!}{%
\includegraphics{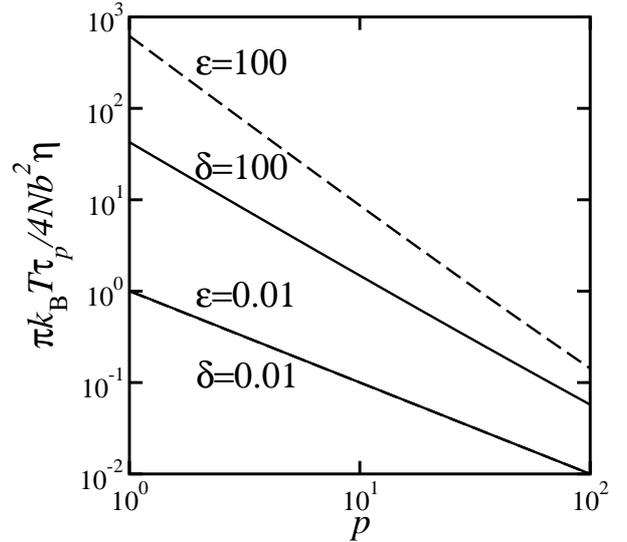}
}
\caption{Scaled relaxation time $\pi k_{\rm B}T \tau_p /4 N b^2 \eta$ 
as a function of $p$ for different values of 
$\delta = 0.01, 100$ (solid lines) and $\varepsilon = 0.01, 100$ 
(dashed lines).
The two curves for $\delta=\varepsilon=0.01$ overlap each other and 
cannot be distinguished.
}
\label{fig:tauvsp}
\end{center}
\end{figure}

In fig.~\ref{fig:taup}, the scaled relaxation time eq.~(\ref{eqn:tausd})
is plotted as a function of $\delta$ for $p=1$ and $10$ as solid lines.
For small $\delta$, the relaxation time is independent of the 
polymer size, which is consistent with eq.~(\ref{eqn:tausdsmalld}).  
The relaxation time increases through a crossover regime towards a 
linear behavior as given by eq.~(\ref{eqn:tausdlarged}).
Such a crossover occurs around the region where the polymer
size $R_{\rm g}$ is comparable to the SD hydrodynamic screening 
length $\nu^{-1}$, i.e., $\delta \sim 1$.
In the limit of large $\delta$, the $p$-dependence of the relaxation 
time is analogous to that obtained from the Zimm model~\cite{doi-edwards}.
The solid lines in fig.~\ref{fig:tauvsp} show the relaxation time as a 
function of the Rouse normal mode $p$ for $\delta=0.01$ and $100$.
These lines have slopes $-1$ and $-3/2$ for $\delta=0.01$
and $100$, respectively. 
These results indicate the different mode dependencies in the two limiting 
polymer sizes.

%%%%%%%%%%%%%%%%%%%%%%%%%%%%%%%%%%%%%%%%%%%%%%%%%%%%
\subsection{Diffusion coefficient}

By substituting eq.~(\ref{eqn:sdmobility}) into eq.~(\ref{eqn:genD}),
the diffusion coefficient of the polymer can be obtained as
\begin{align}
D^{\rm SD}
&=
\frac{k_{\rm B}T}{4 \pi \eta}
\frac{1}{\delta^4}
\left[
( \pi \, {\rm erfi}(\delta) - {\rm Ei}(\delta^2) )
\exp(-\delta^2) 
+ \frac{4\sqrt{\pi}}{3}\delta^3
\right. \nonumber\\
&
+ \delta^2 
- (\ln\delta^2+\gamma) (\delta^2-1) 
- 2\sqrt{\pi} \delta
\bigg],
\label{eqn:Dsd}
\end{align}
where $\gamma = 0.5772\cdots$ is Euler's constant.
This expression for the polymer diffusion coefficient is valid for 
all the ranges of $\delta$.
Equation (\ref{eqn:Dsd}) is one of the main results of this paper.

We now discuss the asymptotic limits of eq.~(\ref{eqn:Dsd}) 
for small and large polymer sizes.  
When $\delta \ll 1$, it reduces to
\begin{equation}
D^{\rm SD} \approx 
\frac{k_{\rm B}T}{4 \pi \eta}
\left( -\ln \delta- \frac{\gamma}{2} + 
\frac{3}{4}\right).
\label{eqn:Dsdsmalld}
\end{equation}
Such a logarithmic behavior is consistent with that of an object 
in a pure 2D system~\cite{saffman-75,saffman-76}.
In the opposite limit of $\delta \gg 1$, we have
\begin{equation}
D^{\rm SD} \approx
\frac{k_{\rm B}T}{4 \pi \eta} \frac{4\sqrt{\pi}}{3\delta}
=\frac{k_{\rm B}T}{6 \sqrt{\pi} \eta_{\rm s} R_{\rm g}}.
\label{eqn:Dsdlarged}
\end{equation}
Similar to eq.~(\ref{eqn:tausdlarged}), this expression depends only 
on $\eta_{\rm s}$.
The obtained $1/\eta_{\rm s}R_{\rm g}$-dependence is analogous 
to that of an object moving in 3D fluid as well as the result by 
Hughes {\it et al.}~\cite{hughes-81}.

\begin{figure}
\begin{center}
\resizebox{0.9\columnwidth}{!}{%
\includegraphics{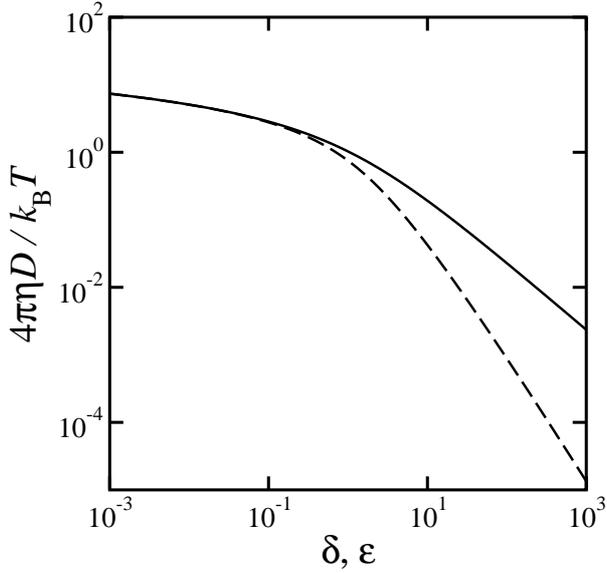}
}
\caption{Dimensionless diffusion coefficient $4 \pi \eta D/k_{\rm B}T$
as a function of $\delta = R_{\rm g} \nu$ (solid line) 
or $\varepsilon = R_{\rm g} \kappa$ (dashed line).}
\label{fig:D}
\end{center}
\end{figure}

In fig.~\ref{fig:D}, we plot the diffusion coefficient $D^{\rm SD}$
as a function of $\delta$ (solid curve).
With the increase in the polymer size, there is a crossover from 
logarithmic to algebraic decay indicated by eqs.~(\ref{eqn:Dsdsmalld}) 
and (\ref{eqn:Dsdlarged}), respectively.
The change in the behavior of a domain of the diffusion coefficient 
with the addition of solvent has also been shown through recent dissipative
particle dynamics simulations on pure 2D and quasi-2D systems~\cite{sanoop-bulk-10}.

%%%%%%%%%%%%%%%%%%%%%%%%%%%%%%%%%%%%%%%%%%%%%%%%%%%%
\subsection{Dynamic structure factor}

The last quantity calculated for the free membrane geometry is the dynamical 
structure factor $S^{\rm SD}({\bf k},t)$ defined in 
eq.~(\ref{eq:skt}).
Since the full expression is rather complicated, we derive 
several analytical expressions for the limiting cases.
For $R_{\rm g} k \ll 1$, we have  
\begin{equation}
S^{\rm SD}({\bf k},t) \approx N \exp (-k^2 D^{\rm SD} t),
\end{equation}
where $D^{\rm SD}$ is given by eq.~(\ref{eqn:Dsd}).
This is reasonable because only the center of mass motion of 
the polymer is captured in the small angle regime.

For $R_{\rm g} k\gg 1$, on the other hand, we only consider the 
time region $t \ll \tau_p^{\rm SD}$ because 
$S^{\rm SD}({\bf k},t)$ becomes very small for 
$t \gg \tau_p^{\rm SD}$.
In this case, we have 
\begin{align}
& S^{\rm SD}({\bf k},t) \approx
\frac{1}{N} \sum_{n,m}
\exp \Big[ -\frac{1}{4} |n-m|b^2k^2 
- \frac{Nb^2k^2}{\pi^2} 
\nonumber \\ 
& \times \sum_{p=1}^\infty
\frac{1}{p^2}
\cos\left( \frac{p \pi n}{N}\right)
\cos\left( \frac{p \pi m}{N}\right)
[ 1 - \exp ( -t/\tau_p^{\rm SD})]
\Big]
\nonumber\\
&= \frac{8}{b^2k^2}
\int_0^\infty {\rm d}u \, 
\exp(-u- k^2 I_1(u)),
\label{eqn:Ssdlarge}
\end{align}
with 
\begin{align}
I_1(u) = \frac{ Nb^2}{2\pi^2}\int_0^\infty {\rm d}p \, 
\frac{1}{p^2}
\cos\left(\frac{4 \pi p u}{N b^2k^2 }\right)
[ 1 - \exp(-t/\tau_p^{\rm SD})].
\end{align}
When $\delta \gg 1$, we can use the limiting expression for 
$\tau_p^{\rm SD}$ as obtained in eq.~(\ref{eqn:tausdlarged}).
In this case, the above expression becomes 
\begin{align}
& S^{\rm SD}({\bf k},t) 
\nonumber\\
&\approx \frac{8}{b^2 k^2} \int_0^\infty {\rm d}u \,
\exp \left[
- u- (\Gamma^{\rm SD} t)^{2/3} w_1( (\Gamma^{\rm SD} t)^{-2/3} u)
\right],
\end{align}
with the decay rate
\begin{equation}
\Gamma^{\rm SD} = \frac{ k_{\rm B}T \sqrt{N} b k^3}
{16 \eta \delta}
=\frac{k_{\rm B}T k^3}{8 \eta \nu},
\label{eq:sddecayrate}
\end{equation}
and 
\begin{equation}
w_1(u) = \frac{2}{\pi} 
\int_0^\infty {\rm d}x \,
\frac{\cos(xu)}{x^2}
[1-\exp (-x^{3/2}/\sqrt{2})].
\end{equation}
Notice the $k^3$-dependence of the decay rate.
For $\Gamma^{\rm SD}t \gg 1$, the above expression is further 
simplified to  
\begin{equation}
S^{\rm SD}({\bf k},t) \approx S^{\rm SD}({\bf k},0)
\exp( -1.35 (\Gamma^{\rm SD} t)^{2/3}),
\label{eqn:SDstretch}
\end{equation}
since $w_1(0) = \Gamma(1/3) \approx 1.35$.
This expression is valid in the limit of large polymer sizes 
such that $R_{\rm g} k \gg 1$ and  $\delta \gg 1$.
Large wave vectors (probing the internal motion of the polymer)
and hydrodynamic screening lengths (high solvent viscosity) 
will also lead to the same expression.
The applicable time window for eq.~(\ref{eqn:SDstretch}) is 
$1/\Gamma^{\rm SD}\ll t \ll \tau_p^{\rm SD}$.
These expressions for large $\delta$ are analogous to that obtained 
from the Zimm model~\cite{doi-edwards} in the $R_{\rm g} k \gg 1$ regime
for polymer in 3D.

%%%%%%%%%%%%%%%%%%%%%%%%%%%%%%%%%%%%%%%%%%%%%%%%%%%%
\section{Polymer dynamics: confined membrane case}
\label{polyES}
%%%%%%%%%%%%%%%%%%%%%%%%%%%%%%%%%%%%%%%%%%%%%%%%%%%%

When the thickness of the solvent layers is very small, 
the membrane is now almost confined by the two walls.
However, there is a thin lubricating layer between the membrane 
and the walls so that $h \neq 0$. 
In this case, we use the ES mobility tensor given by   
eq.~(\ref{eqn:esoseen}). 
Several quantities for this case are obtained below.

%%%%%%%%%%%%%%%%%%%%%%%%%%%%%%%%%%%%%%%%%%%%%%%%%%%%
\subsection{Mobility tensor}

By using eq.~(\ref{eqn:esoseen}), the preaveraged mobility tensor 
is calculated from
\begin{align}
\langle {\bf G}^{\rm ES}_{nm}\rangle
= \left\langle
\int \frac{ {\rm d}^2 k}{(2 \pi)^2}
\frac{ {\mathbf I} -  \hat {\bf k} \hat {\bf k}}
{\eta(k^2 + \kappa^2)}
\exp [i {\bf k} \cdot ({\bf R}_n - {\bf R}_m )]
\right\rangle.
\end{align}
This results in $\langle {\bf G}^{\rm ES}_{nm}\rangle 
=g^{\rm ES}(n-m) {\bf I}$ with
\begin{align}
& g^{\rm ES}(n-m) \nonumber \\
& = \frac{1}{4 \pi \eta} \int_0^\infty {\rm d}k \,
\frac{k}{ k^2 + \kappa^2}
\exp\left( -\frac{1}{4} b^2 k^2 \vert n-m \vert \right)
\nonumber\\
&=-\frac{1}{8\pi\eta}
\exp\left(  \frac{1}{4} b^2 \kappa^2 \vert n-m \vert \right)
{\rm Ei}\left(- \frac{1}{4}b^2 \kappa^2 \vert n-m \vert \right),
\label{eqn:esmobility}
\end{align}
which was previously derived in ref.~\cite{komura-95}.

%%%%%%%%%%%%%%%%%%%%%%%%%%%%%%%%%%%%%%%%%%%%%%%%%%%%
\subsection{Relaxation time}

The polymer relaxation time can be obtained by substituting 
eq.~(\ref{eqn:esmobility}) into eq.~(\ref{eqn:gengpq}).
Then we have 
\begin{align}
g^{\rm ES}_{p}  
&=
\frac{1}{ \pi \eta Nb^2}
\int_0^\infty {\rm d}k \,
\frac{k^3}
{(k^2 + \kappa^2)
[ k^4 + \left( 4 \pi p/N b^2 \right)^2 ]}
\nonumber\\
&=
\frac{1}{16 \pi \eta} 
\frac{ 
\pi^2 p + 2 \varepsilon^2 \ln(\varepsilon^2 /(\pi p))
}
{\pi^2 p^2 + \varepsilon^4 }.
\end{align}
In the above, we have defined the dimensionless polymer size as 
$\varepsilon \equiv \sqrt{N}b \kappa/2 = R_{\rm g} \kappa$ which should
be distinguished from $\delta$ in the previous section.
Then the relaxation time can be written as 
\begin{equation}
\tau^{\rm ES}_p = \frac{4 N b^2 \eta}{\pi k_{\rm B} T} 
\frac
{\pi^2 p^2 + \varepsilon^4 }
{p^2[\pi^2 p + 2 \varepsilon^2 \ln(\varepsilon^2 /(\pi p))]}.
\label{eqn:taues}
\end{equation}
In the limit of $\varepsilon \ll 1$, it reduces to
\begin{equation}
\tau_p^{\rm ES} \approx \frac{4 N b^2 \eta}{\pi k_{\rm B} T} 
\frac{1}{p}, 
\label{eqn:tauessmalle}
\end{equation}
which coincides with eq.~(\ref{eqn:tausdsmalld}).
In the opposite limit of $\varepsilon \gg 1$, one gets
\begin{equation}
\tau_p^{\rm ES}\approx
\frac{4 N b^2 \eta}{\pi k_{\rm B}T}
\frac{\varepsilon^2}{2p^2 \ln(\varepsilon^2/(\pi p))}.
\label{eqn:taueslargee}
\end{equation}
In fig.~\ref{fig:taup}, we plot $\tau_p^{\rm ES}$ as a function 
of $\varepsilon$ for $p=1$ and $10$ in dashed lines. 
The algebraic $\varepsilon^2$-dependence in eq.~(\ref{eqn:taueslargee})    
is seen for large $\varepsilon$.
The dashed lines in fig.~\ref{fig:tauvsp} are the plots of 
$\tau_p^{\rm ES}$ as a function of $p$ for $\varepsilon=0.01$ and 
$100$. 
For $\varepsilon=100$, the slope $-2$ is consistent with
eq.~(\ref{eqn:taueslargee}) neglecting the logarithmic correction.    
Notice that this $p$-dependence is in contrast to that 
for the free membrane case given in eq.~(\ref{eqn:tausdlarged}).

%%%%%%%%%%%%%%%%%%%%%%%%%%%%%%%%%%%%%%%%%%%%%%%%%%%%
\subsection{Diffusion coefficient}

With the use of eq.~(\ref{eqn:esmobility}), the diffusion coefficient 
for the confined membrane geometry is written as  
\begin{align}
D^{\rm ES}& =
\frac{k_{\rm B}T}{4 \pi \eta}
\frac{1}{\varepsilon^4}
[(1+\varepsilon^2)(2 \ln \varepsilon + \gamma)
- \varepsilon^2 
\nonumber \\
& -\exp(\varepsilon^2){\rm Ei}(-\varepsilon^2)].
\label{eqn:Des}
\end{align}
This equation was also obtained before~\cite{komura-95}.
The limiting expression for $\varepsilon \ll 1$ is 
\begin{equation}
D^{\rm ES} \approx 
\frac{k_{\rm B}T}{4 \pi \eta}
\left(-\ln \varepsilon- \frac{\gamma}{2} 
+ \frac{3}{4} \right),
\label{eqn:Dessmalle}
\end{equation}
which coincides with eq.~(\ref{eqn:Dsdsmalld}) as long as 
$\varepsilon$ is replaced by $\delta$. 
When $\varepsilon \gg 1$, eq.~(\ref{eqn:Des}) reduces to
\begin{equation}
D^{\rm ES}\approx 
\frac{k_{\rm B}T}{4 \pi \eta}
\frac{1}{\varepsilon^2}
=\frac{k_{\rm B}T h}{8 \pi \eta_{\rm s} R_{\rm g}^2}.
\label{eqn:Deslargee}
\end{equation}
This $1/R_{\rm g}^2$-dependence is a characteristic of a system in 
which there is momentum loss from the membrane to the surrounding
environment~\cite{diamant-09b}.
An intuitive understanding of the large size behaviors of the 
diffusion coefficient in terms of the conservation principles 
will be described in sect.~\ref{discussion}.
The dashed line in fig.~\ref{fig:D} shows the plot of $D^{\rm ES}$ as 
a function of $\varepsilon$, showing the logarithmic and algebraic 
behaviors as derived above.

According to the definition of $\delta$ and $\varepsilon$,
one obtains $\varepsilon=\delta/\sqrt{\nu h}$.
This correspondence leads to a rescaling of the $\varepsilon$-axis in
figs.~\ref{fig:taup} and~\ref{fig:D}.
Here the value of $\sqrt{\nu h}$ cannot be taken arbitrarily
since the condition for the confined membrane is given by 
$\nu h \ll 1$ as explained in sect.~\ref{membrane}.

%%%%%%%%%%%%%%%%%%%%%%%%%%%%%%%%%%%%%%%%%%%%%%%%%%%%
\subsection{Dynamic structure factor}

The dynamic structure factor can be calculated in the same manner
as before. 
For $R_{\rm g}k \ll 1$, we have 
\begin{equation}
S^{\rm ES}({\bf k},t) \approx 
N \exp(-k^2 D^{\rm ES} t).
\end{equation}
For $R_{\rm g} k \gg 1$ and $t \ll \tau_p^{\rm ES}$, we get 
\begin{align}
S^{\rm ES}({\bf k},t) \approx
\frac{8}{b^2k^2}
\int_0^\infty {\rm d}u \, 
\exp (-u - k^2 I_2(u)),
\label{eqn:Seslarge}
\end{align}
with 
\begin{align}
I_2(u)=  \frac{ Nb^2}{2\pi^2}\int_0^\infty {\rm d}p \,
\frac{1}{p^2}
\cos\left(\frac{4 \pi p u}{N b^2k^2 }\right)
[ 1 - \exp(-t/\tau_p^{\rm ES})].
\end{align}
Considering $\varepsilon \gg 1$ and neglecting the logarithmic 
dependence in eq.~(\ref{eqn:taueslargee}), the above expression
becomes 
\begin{align}
&S^{\rm ES}({\bf k},t) 
\nonumber \\
&\approx \frac{8}{b^2 k^2} 
\int_0^\infty {\rm d}u \, 
\exp \left[
- u- (\Gamma^{\rm ES} t)^{1/2} w_2((\Gamma^{\rm ES} t)^{-1/2}u)
\right],
\end{align}
with the decay rate 
\begin{equation}
\Gamma^{\rm ES} = \frac{k_{\rm B} T N b^2 k^4}
{32 \pi \eta \varepsilon^2}
=\frac{k_{\rm B} T k^4}{8\pi \eta \kappa^2},
\label{eq:esdecayrate}
\end{equation}
and
\begin{align}
w_2(u) = \frac{2}{\pi}
\int_0^\infty {\rm d}x \,
\frac{\cos(xu)}{x^2} [1-\exp(-x^2)]. 
\end{align}
Note that $\Gamma^{\rm ES}$ is proportional to $k^4$
as for the Rouse model. 
For $\Gamma^{\rm ES}t \gg 1$, it can be further simplified to  
\begin{align}
S^{\rm ES}({\bf k},t)\approx S^{\rm ES}({\bf k},0)
\exp( - 1.13(\Gamma^{\rm ES} t)^{1/2} ),
\end{align}
since $w_2(0)=2/\sqrt{\pi}\approx 1.13$.
This expression is valid when $R_{\rm g} k \gg 1$ 
and $\varepsilon \gg 1$.
The relevant time interval for this expression is
$1/\Gamma^{\rm ES} \ll t \ll \tau_p^{\rm ES}$.

%%%%%%%%%%%%%%%%%%%%%%%%%%%%%%%%%%%%%%%%%%%%%%%%%%%%
\section{Excluded volume effects}
\label{excV}
%%%%%%%%%%%%%%%%%%%%%%%%%%%%%%%%%%%%%%%%%%%%%%%%%%%%

So far, we have treated only a 2D Gaussian polymer chain.
In this section, we briefly discuss the effects of excluded 
volume on the dynamical quantities.
Even if the polymer is not a Gaussian chain, one can still
use the preaveraging approximation for the mobility tensor.
For the free membrane case, one can generally show that 
\begin{align}
g^{\rm SD}(n-m)=\frac{1}{8\eta}
\langle {\bf H}_0( \nu r_{nm} ) - Y_0( \nu r_{nm} )\rangle,
\label{eq:exvSD}
\end{align}
where $r_{nm}= \vert {\bf R}_n - {\bf R}_m \vert$
(see appendix B).
Then the diffusion coefficient can be expressed as 
\begin{equation}
D^{\rm SD}= 
\frac{k_{\rm B}T}{8\eta}
\int_0^N \frac{{\rm d} n}{N}
\int_0^N \frac{{\rm d} m}{N}
\langle {\bf H}_0( \nu r_{nm}) - Y_0( \nu r_{nm}  )\rangle.
\label{eqn:kirkSD}
\end{equation}
Similarly for the confined membrane case, we find 
\begin{equation}
D^{\rm ES}= 
\frac{k_{\rm B}T}{4 \pi \eta}
\int_0^N \frac{{\rm d} n}{N}
\int_0^N \frac{{\rm d} m}{N}
\langle K_0(  \kappa r_{nm} )\rangle,
\label{eqn:kirkES}
\end{equation}
(see eq.~(\ref{eqn:es2AB})).
These expressions are the 2D analog of the Kirkwood 
formula~\cite{doi-edwards}.
It should be emphasized that they are rigorous even in the presence 
of excluded volume effect.

For excluded volume chains, however, the appropriate 
equilibrium distribution function ${\rm \Psi}(\{{\bf R}_n\}) $
needed to calculate averages is not known~\cite{doi-edwards}.
Therefore, we cannot obtain rigorous forms of diffusion coefficient. 
Instead, we shall make use of scaling arguments to infer the effects 
of excluded volume interactions. 
Here, we limit our discussion to small and large polymer sizes. 
A simple argument is that the excluded volume effects lead to a 
rescaled polymer radius of gyration $R_{\rm g}$. 
Within the Flory theory, the radius of gyration scales as 
$R_{\rm g} \sim b N^{\nu_{\rm F}}$ using the Flory exponent $\nu_{\rm F}$.
Up to a numerical factor, the diffusion coefficient in the limiting 
cases will still show the same size dependence as in 
eqs.~(\ref{eqn:Dsdsmalld}), (\ref{eqn:Dsdlarged}), (\ref{eqn:Dessmalle})
and (\ref{eqn:Deslargee}) in which $R_{\rm g}$ is now replaced 
with that of excluded volume chains. 
In other words, we replace $N$ with $N^{2\nu_{\rm F}}$ in these equations.
On the other hand, the relaxation time in the presence of the 
excluded volume effects can be obtained by replacing $N/p$ with 
$(N/p)^{2\nu_{\rm F}}$ in the Gaussian chain cases.
In the small size limits, i.e, $\delta, \varepsilon \ll 1$, we have   
\begin{equation}
\tau^{\rm SD} 
\sim \frac{\eta b^2}{k_{\rm B} T} 
\left(\frac{N}{p} \right)^{2 \nu_{\rm F}},
\end{equation}
and 
\begin{equation}
\tau^{\rm ES} 
\sim \frac{\eta b^2}{k_{\rm B} T} 
\left(\frac{N}{p} \right)^{2 \nu_{\rm F}}.
\end{equation}
In the opposite of large polymer size limits, i.e., $\delta,\varepsilon \gg 1$, 
we get
\begin{equation}
\tau^{\rm SD} 
\sim \frac{\eta b^3 \nu}{k_{\rm B} T} 
\left(\frac{N}{p} \right)^{3 \nu_{\rm F}},
\end{equation}
and 
\begin{equation}
\tau^{\rm ES} 
\sim \frac{\eta b^4 \kappa^2}{k_{\rm B} T}
\left(\frac{N}{p} \right)^{4 \nu_{\rm F}}.
\end{equation}
These are the scaling predictions.
Notice that the above results can also be obtained by using the 
relation $\tau \sim R_{\rm g}^2/D$ where $R_{\rm g}$ includes 
the excluded volume effect~\cite{rubinstein-colby}.

Table~\ref{tab:1} shows the comparison between a Gaussian chain polymer 
($\nu_{\rm F} = 1/2$) and a chain with excluded volume 
interactions ($\nu_{\rm F}=3/4$) in 2D.
The former Gaussian case recovers all the relations in the previous
sections (see eqs.~(\ref{eqn:tausdsmalld}), (\ref{eqn:tauessmalle}),
(\ref{eqn:tausdlarged}), (\ref{eqn:taueslargee})).
For the chain with $\nu_{\rm F}=3/4$, the relaxation time shows 
a  $p^{-3/2}$-dependence for both the free and confined membrane 
cases when $\delta,\varepsilon \ll 1$.
This is consistent with the result in ref.~\cite{muthukumar-85}.
For $\delta \gg 1$ and $\varepsilon \gg 1$, we have 
$\tau^{\rm SD} \sim p^{-9/4}$ and $\tau^{\rm ES} \sim p^{-3}$,
respectively.
These exponents are unique for polymers confined in a membrane.

\begin{table}
\caption{Comparison of the relaxation times between a Gaussian
polymer chain and a polymer with excluded volume interactions.}
\label{tab:1}      
\begin{tabular}{ccc}
\hline\noalign{\smallskip}
limits & ideal chain ($\nu_{\rm F} = 1/2$)  & 
real chain ($\nu_{\rm F} = 3/4$) \\
\noalign{\smallskip}\hline\noalign{\smallskip}
$\delta \ll 1$ & $\tau^{\rm SD} \sim p^{-1}$ & 
$\tau^{\rm SD} \sim p^{-3/2}$ \\
$\varepsilon \ll 1$ & $\tau^{\rm ES} \sim p^{-1}$   
& $\tau^{\rm ES} \sim p^{-3/2}$ \\
$\delta \gg 1$   & $\tau^{\rm SD} \sim p^{-3/2}$ & 
$\tau^{\rm SD} \sim p^{-9/4}$ \\
$\varepsilon \gg 1$ & $\tau^{\rm ES} \sim p^{-2}$   
& $\tau^{\rm ES} \sim p^{-3}$ \\
\noalign{\smallskip}\hline
\end{tabular}
\end{table}

%%%%%%%%%%%%%%%%%%%%%%%%%%%%%%%%%%%%%%%%%%%%%%%%%%%%
\section{Discussion}
\label{discussion}
%%%%%%%%%%%%%%%%%%%%%%%%%%%%%%%%%%%%%%%%%%%%%%%%%%%%

In this paper, we have investigated the dynamics of a Gaussian 
polymer chain confined in a liquid membrane taking into account 
the surrounding environment. 
For the most general geometry with the membrane, solvent and walls
(see fig.~\ref{fig:membwall}), we have derived the mobility tensor 
in eq.~(\ref{eqn:genoseen}). 
We obtained the analytical expressions for the relaxation time, 
diffusion coefficient and dynamic structure factor of a 
Gaussian chain for the two limiting cases of free and 
confined membranes. 
These quantities were calculated within the preaveraging approximation
of the mobility tensor in order to avoid the non-linearity introduced 
by the hydrodynamic coupling. 
We shall summarize and discuss the results obtained in this paper.

Our theoretical analysis relies on the preaveraging approximation, which is
used to decouple the polymer fluctuations from hydrodynamics. 
The validity of this approximation has been studied in some detail for 
polymers in a 3D bulk fluid, where the preaveraging approximation (Zimm model) 
yields results which are not very different from more sophisticated
calculations~\cite{doi-edwards}.
The response of a dilute polymer solution to an external field has also been
experimentally verified to follow the predictions of the Zimm 
model~\cite{johnson-70}.
Deviations from the Zimm model have been attributed to non-Gaussian 
distributions, the slowness to reach asymptotic behaviors or the effect 
of hydrodynamic fluctuations, although a clear conclusion has not been 
reached~\cite{stockmayer-84}.
Up to now, there have not been any experimental studies of the dynamics of
polymers confined to membranes, to the best of our knowledge. 
Experiments using a combination of Langmuir trough as well as light
scattering techniques should be able to test our predictions. 
The success of preaveraging in three-dimensional polymer solutions encourages 
us to expect the same to hold for quasi-2D systems.

We first compare the relaxation times between the two cases.
As seen from eqs.~(\ref{eqn:tausdsmalld}) and (\ref{eqn:tauessmalle})
for small polymer sizes, i.e., $\delta, \varepsilon \ll 1$, both 
$\tau_p^{\rm SD}$ and $\tau_p^{\rm ES}$ show exactly the same
mode dependence. 
This is due to the fact that for polymer sizes smaller than than 
the hydrodynamic screening lengths ($\nu^{-1}$ or $\kappa^{-1}$),
the outer environment does not affect the polymer dynamics. 
The behavior $\tau_p^{\rm SD} \sim p^{-3/2}$ for $\delta \gg 1$
(see eq.~(\ref{eqn:tausdlarged})) is analogous to the Zimm 
relaxation time of a Gaussian polymer in a 3D solvent. 
On the other hand, the dependence $\tau_p^{\rm ES} \sim p^{-2}$
for $\varepsilon \gg 1$ (see eq.~(\ref{eqn:taueslargee})) is similar 
to the Rouse relaxation time in 3D.

One of the important results of this paper is the derivation of 
eq.~(\ref{eqn:Dsd}) which provides the diffusion coefficient valid 
for all size ranges in the free membrane case.
According to eqs.~(\ref{eqn:Dsdsmalld}) and (\ref{eqn:Dessmalle}),
the diffusion coefficients $D^{\rm SD}$ and $D^{\rm ES}$ show the 
same logarithmic dependence when $\delta,\varepsilon \ll 1$.
Again this occurs when the polymer size is smaller 
than $\nu^{-1}$ or $\kappa^{-1}$.
A crossover from logarithmic to algebraic behavior takes place 
when the polymer size $R_{\rm g}$ is comparable to these hydrodynamic 
screening lengths.   
Thus $\nu^{-1}$ or $\kappa^{-1}$ determines the length scale at which
the outer environment surrounding the membrane becomes important.
For a pure 2D system, the screening length is infinite, which causes
the Stokes paradox.

When the polymer size becomes much larger than the hydrodynamic
screening lengths, the interactions are no longer only through the 
membrane.
In the free membrane case, the outer fluid plays the role of the mediator
of the hydrodynamic interactions.
This attributes a 3D nature to the polymer dynamics and hence 
the scaling $D^{\rm SD} \sim 1/\eta_{\rm s}R_{\rm g}$ is recovered (see 
eq.~(\ref{eqn:Dsdlarged})).
In the confined membrane case, the presence of the walls takes away 
the momentum from the membrane.
Owing to the stick boundary condition, a linear shear velocity 
gradient is set up in the intervening fluid layer between the 
membrane and the walls leading to a momentum leakage from the 
membrane. 
As given in eq.~(\ref{eqn:Deslargee}), the resultant diffusion 
coefficient behaves as $D^{\rm ES} \sim 1/R_{\rm g}^2$.
These asymptotic behaviors are consistent with the results of 
correlated diffusion obtained in ref.~\cite{oppenheimer-10}.

The different large size behaviors of the diffusion coefficient
can be better understood by focusing on the 
mobility tensors.
It is essentially described as a consequence of the conservation of 
mass and momentum principles~\cite{diamant-09b}. 
The mobility tensor ${\bf G}({\bf r})$ acts as a Green's function 
relating a source of disturbance at the origin to the velocity at 
a point ${\bf r}$.
At sufficiently large distances, the source of disturbance in the 
fluid can be thought of as a force monopole which introduces 
a source for momentum in the fluid.
A moving object also causes a perturbation in the mass density and 
this can be regarded as a mass dipole (a source and a sink of mass 
density).  
We now apply the conservations of momentum and mass to the 3D and 2D 
cases separately.

For the 3D case, the presence of a force monopole at the origin  should
cause the momentum flux or stress $\sigma$ decay as $1/r^2$ in 
order to conserve the total momentum ($r^2$ being proportional to 
the area of a sphere surrounding the force monopole). 
Since the shear stress is related to the fluid velocity through 
$\sigma \sim \eta_{\rm s} v/r$, we have $v \sim 1/\eta_{\rm s} r$.
This implies that the mobility tensor should also scale as 
$1/ \eta_{\rm s}r$.
Concerning the mass conservation principle, a mass monopole would 
create a flow velocity that decays as $1/r^2$.
Since we have a mass dipole, the resulting velocity now decays
as $1/r^3$.
Comparing these two effects, the contribution to the velocity due 
to momentum conservation (which varies as $1/r$) always dominates 
at large distances in 3D. 
This essentially explains the behavior 
$D^{\rm SD} \sim 1/\eta_{\rm s}R_{\rm g}$.

Let us now consider the 2D case in which the membrane is in contact 
with the walls leading to a loss of momentum from the membrane.
This implies that the momentum is not conserved, and the only
contribution to the velocity is from the mass conservation.
In 2D, a mass monopole will create a velocity which decays as $1/r$
($r$ being the perimeter of a circle surrounding the mass monopole).
Hence the velocity and the mobility tensor due to a mass dipole 
decays as $1/r^2$. 
This explains the scaling $D^{\rm ES} \sim 1/R_{\rm g}^2$. 
These two different behaviors of the mobility tensors are reflected 
in the diffusion coefficients for large size polymers.
Incidentally, for the pure 2D case, the stress decays as $1/r$ due 
to the momentum conservation.
Since the stress scales as $\sigma \sim \eta v/r$, we have 
$v \sim 1/\eta$.   
This explains the logarithmic size dependence of the diffusion 
coefficient.

The dynamic structure factor is a quantity readily accessible 
through scattering experiments.
For small wave numbers $R_{\rm g} k \ll 1$, only the center of mass 
motion of the polymer can be captured.
When $R_{\rm g} k \gg1$ and $t$ is much less than the relaxation times, 
the screening lengths become important.
In the limit of 
$\delta \gg 1$, $S^{\rm SD}({\bf k},t)$ shows a 
stretched exponential decay with an exponent $2/3$, and so does 
$S^{\rm ES}({\bf k},t)$ for $\varepsilon \gg 1$ with an exponent $1/2$.
Again the role of the outer environment is reflected in the decay 
rates $\Gamma^{\rm SD}$ and $\Gamma^{\rm ES}$ given by 
eqs.~(\ref{eq:sddecayrate}) and (\ref{eq:esdecayrate}), respectively.
In the free membrane case, the dependence $\Gamma^{\rm SD} \sim k^3$ resembles 
that of the Zimm model in 3D. 
For confined membranes on the other hand the behavior $\Gamma^{\rm ES} \sim 
k^4$ is analogous to that obtained from the Rouse model.

The dynamics of a hydrophobic polymer embedded in a 2D membrane 
was previously investigated by Muthukumar both for a Gaussian 
polymer and a polymer with excluded volume 
effects~\cite{muthukumar-85}.
In his treatment, the membrane was regarded as an isolated 
entity without any couplings to the outer environment. 
The membrane 
2D nature was taken into account through an anisotropic viscosity.
It was shown that the longest relaxation time is proportional 
to $p^{-1}$ for a Gaussian chain. 
This agrees with our limiting expressions of the relaxation times 
in eqs.~(\ref{eqn:tausdsmalld}) and~(\ref{eqn:tauessmalle}) obtained
for small $\delta$ and $\varepsilon$, respectively.
In ref.~\cite{muthukumar-85}, the excluded volume effects were 
taken into account by a mode-dependent polymer blob size.
In this case, he showed that the relaxation time scales as $p^{-3/2}$.
As discussed in sect.~\ref{excV}, we find the mode dependencies of 
the relaxation times are altered in the presence of excluded volume 
interactions.
In the small size limit, we indeed recover the scaling $p^{-3/2}$ 
for the free and confined membrane cases (see table~\ref{tab:1}). 
Notice again that the polymer dynamics in this limit remains 
unaffected by the outer environment. 
For large polymer sizes, we obtain either $\tau^{\rm SD} \sim p^{-9/4}$ 
or $\tau^{\rm ES} \sim p^{-3}$ for real polymer chains.

A related situation to a polymer in a confined membrane 
is a dilute polymer solution trapped in a slit-like geometry whose 
width is much smaller than the polymer blob size.
Using scaling arguments, Brochard calculated the polymer relaxation time 
scales as $p^{-5/2}$~\cite{broch-77}.
An experimental realization of such a geometry was done by Lin {\it et al.}
who confined dilute DNA solution in quasi-2D 110 nm wide slits~\cite{lin-07}. 
The relaxation times measured in this case was found to scale as 
$p^{-2.2}$.
However, it should be emphasized that this scenario is different from the
model discussed in this article.
In our model, the polymer chain is strictly confined to the 2D plane of 
the membrane which itself is embedded in a 3D bulk fluid.

At this stage, a rough estimate of the screening lengths would 
be useful. 
As reported in ref.~\cite{peters-82}, 
the membrane viscosity of
dimyristoylphosphatidylcholine bilayers at $32^\circ {\rm C}$ 
(rounded to the nearest order) 
is approximately $0.1$ Ns/m$^2$ and the viscosity of water
is $\eta_{\rm s} \approx 10^{-3}$ Ns/m$^2$.
For supported membranes we can approximate the height of the intervening
solvent region to be $h \approx 10^{-8}$ m~\cite{kaizuka-04}.
Hence we obtain 
$\nu^{-1} \approx 2.5 \times 10^{-7}$ m  and
$\kappa^{-1} \approx 0.5 \times 10^{-7}$ m.
As described in the Introduction, the experiments with DNA adsorbed on 
supported membranes point towards a Rouse-like 
behavior~\cite{maier-99,maier-00}.
The strong electrostatic attraction between the negatively charged
DNA and the positively charged membrane prevents any out-of-plane
motions of the polymer.
Because the typical length scale of a DNA molecule used in the 
experiments were of several microns, 
the scenario should be close to the case of 
$\varepsilon \gg 1$. 
Hence our result is consistent with the experimental observations 
showing the Rouse-like behavior.
A more recent experiment on DNA adsorption on free standing cationic 
giant unilamellar vesicles showed that the diffusion coefficient of 
DNA molecules lies in the crossover region between the logarithmic 
to the algebraic regimes~\cite{herold-10}.

We finally discuss the mobility tensor for a supported membrane.
In this case, a membrane sits at the bottom of a trough filled 
with a bulk solution having an infinite depth when compared 
to the membrane thickness. 
In fig.~\ref{fig:membwall}, this corresponds to the case where 
$h^+$ is infinitely large while $h^-$ is finite.
Using the general expression of the mobility tensor in 
eq.~(\ref{eqn:genoseen}), we obtain
\begin{equation}
G_{\alpha \beta}[ {\bf k} ] 
= \frac{1}{ \eta k^2 + k[\eta_{\rm s}^+ 
 + \eta_{\rm s}^-\coth(kh^-)]}
\left( \delta_{\alpha\beta} - \frac{k_\alpha k_\beta}{k^2} 
\right).
\end{equation}
If we further assume that $h^-$ is very small, the above equation 
reduces to 
\begin{equation}
G_{\alpha \beta}[ {\bf k} ] 
= \frac{1}{ \eta k^2 + \eta_{\rm s}^+ k
 + (\eta_{\rm s}^-/h^-)}
\left( \delta_{\alpha\beta} - \frac{k_\alpha k_\beta}{k^2} 
\right).
\end{equation}
This mobility tensor for a supported membrane contains two 
screening length scales, i.e.,
$\eta/\eta_{\rm s}^+$ and $(\eta h^-/\eta_{\rm s}^-)^{1/2}$.
The investigation using the above mobility tensor is left as our 
future work.
\\

%%%%%%%%%%%%%%%%%%%%%%%%%%%%%%%%%%%%%%%%%%%%
We thank H. Diamant, Y. Fujitani, M. Imai, T. Kato and 
N. Oppenheimer for useful discussions.
This work was supported by KAKENHI (Grant-in-Aid for Scientific
Research) on Priority Area ``Soft Matter Physics'' and Grant
No.\ 21540420 from the Ministry of Education, Culture, Sports, 
Science and Technology of Japan.

%%%%%%%%%%%%%%%%%%%%%%%%%%%%%%%%%%%%%%%%%%%%
\renewcommand{\theequation}{A.\arabic{equation}}  
\setcounter{equation}{0}  % reset counter     
\section*{Appendix A. Derivation of the general mobility tensor}

\begin{figure}
\begin{center}
\resizebox{0.9\columnwidth}{!}{%
\includegraphics{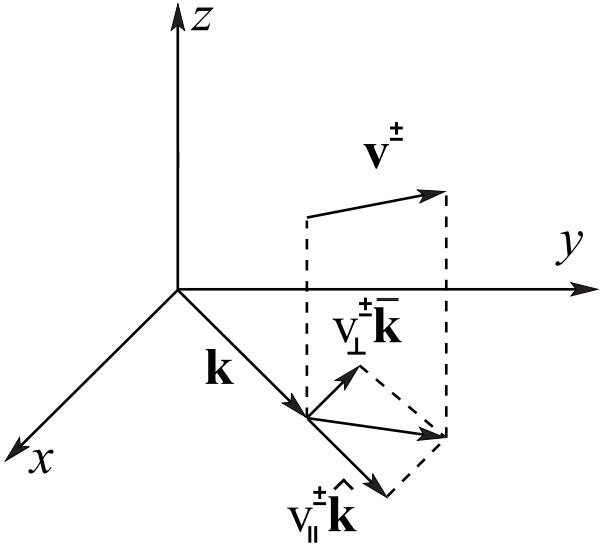}
}
\caption{Projection of ${\bf v}^\pm[{\bf k},z]$ onto the 
$xy$-plane which is decomposed as 
$v_{\parallel}^\pm \hat{{\bf k}} + v_\perp^\pm\bar{{\bf k}}$.}
\label{fig:kperppar}
\end{center}
\end{figure}

Our derivation of the mobility tensor for the membrane closely 
follows that by Inaura and Fujitani~\cite{inaura-08}.
As presented in fig.~\ref{fig:membwall}, we consider a more general 
case where the two walls are located at different distances from the 
membrane, i.e., $h^+ \neq h^-$.   
The membrane is assumed to be impermeable.

Our purpose is to derive the in-plane force ${\bf f}_{\rm s}$ on the 
membrane due to the bulk solvent and walls 
(see eq.~(\ref{eqn:2Dstokes})). 
We first take the Fourier transform of ${\bf v}^\pm({\bf r},z)$ by
\begin{equation}
{\bf v}^\pm[{\bf k},z] = 
\int {\rm d}^2 r \,
{\bf v}^\pm({\bf r},z)
\exp ( -i {\bf k}\cdot {\bf r}),
\label{eqn:FTsolvent}
\end{equation}
where ${\bf r}=(x,y)$ and ${\bf k}=(k_x,k_y)$.
The projection of the vector ${\bf v}^\pm[{\bf k},z]$ on the
$xy$-plane can be expressed as 
$v_{\parallel}^{\pm}[{\bf k},z] \hat{{\bf k}} + 
v_{\perp}^{\pm}[{\bf k},z] \bar{{\bf k}}$ 
where $\hat{{\bf k}} = (k_x/k,k_y/k)$ and
$\bar{{\bf k}} = (-k_y/k,k_x/k)$ with 
$k = \vert {\bf k} \vert$.
The subscripts $\parallel$ and $\perp$ indicate the components 
parallel and perpendicular to $\hat{{\bf k}}$, respectively.
From eq.~(\ref{eqn:3Dstokes}), the vertical component 
$v_{\perp}^{\pm}[{\bf k},z]$ obeys the equation 
\begin{equation}
\left(-k^2 + \frac{\partial^2}{\partial z^2}\right)
v_{\perp}^{\pm}[{\bf k},z]= 0.
\label{eqn:zcompeqn}
\end{equation}
The solution to the above equation can be written as 
\begin{equation}
v_{\perp}^{\pm}[{\bf k},z]
= A_1e^{-kz}+A_2e^{kz},
\end{equation}
with unknown coefficients $A_1$ and $A_2$.
These coefficients are determined by the stick boundary conditions 
imposed at the membrane-solvent and solvent-wall boundaries:
\begin{equation}
v_{\perp}^{\pm}[{\bf k},z] = 
\begin{cases} 
v_{\perp}[{\bf k}]  & \mbox{at}~~ z=0, \\
0 & \mbox{at}~~ z=\pm h^{\pm}, 
\end{cases}
\end{equation}
where $v_\perp[{\bf k}] = {\bf v}[{\bf k}]\cdot \bar{{\bf k}}$ and 
see eq.(\ref{eqn:FT}) for the definition of ${\bf v}[{\bf k}]$.
Thus we have 
\begin{equation}
v_{\perp}^{\pm}[{\bf k},z]
= \mp v_\perp[{\bf k}] 
\frac{\sinh (k(z\mp h^\pm))}{\sinh(kh^\pm)}.
\label{eqn:fluidvel}
\end{equation}

From eq.~(\ref{eqn:fluidvel}), the in-plane force on the membrane
due to the outer solvent and walls becomes
\begin{align}
f_{\rm s\perp}[{\bf k}] 
& = {\bf f}_{\rm s}[{\bf k}] \cdot \bar{{\bf k}}
\nonumber \\
& = \frac{\partial}{\partial z} 
\left( \eta^{+}_{\rm s} v_\perp^{+}[{\bf k},z]
- \eta^{-}_{\rm s} v_\perp^{-}[{\bf k},z] \right)_{z=0}  
\nonumber\\
&= - \eta_{\rm s}^+ v_\perp k \coth(kh^+)
- \eta_{\rm s}^- v_\perp k \coth(kh^-).
\end{align}
Since the incompressibility condition of the membrane fluid
implies $v_{\parallel}[{\bf k}] = 
{\bf v}[{\bf k}] \cdot \hat{{\bf k}}= 0$, 
the perpendicular component of the Fourier transform of 
eq.~(\ref{eqn:2Dstokes}) gives
\begin{equation}
-\eta k^2 v_{\perp}[{\bf k}] +
f_{{\rm s}\perp}[{\bf k}]+
F_{\perp}[{\bf k}]=0, 
\end{equation}
with $F_{\perp}[{\bf k}] = {\bf F}[{\bf k}] \cdot \bar{{\bf k}}$.
Hence the mobility tensor defined by 
${\bf v}[{\bf k}]={\bf G}[{\bf k}] \cdot {\bf F}[{\bf k}]$
is given by eq.~(\ref{eqn:genoseen}).

%%%%%%%%%%%%%%%%%%%%%%%%%%%%%%%%%%%%%%%%%%%%
\renewcommand{\theequation}{B.\arabic{equation}}  
\setcounter{equation}{0}  % reset counter     
\section*{Appendix B. Mobility tensors in real space}

%%%%%%%%%%%%%%%%%%%%%%%%%%%%%%%%%%%%%%%%%%%%
\subsection*{Free membrane case}

The mobility tensor in Fourier space is given by 
eq.~(\ref{eqn:sdoseen}).
The expression for $G^{\rm SD}_{\alpha\beta}({\bf r})$ can
be found by assuming
\begin{equation}
G^{\rm SD}_{\alpha\beta}({\bf r}) = 
B_1 \delta_{\alpha\beta} + B_2 \frac{r_\alpha r_\beta}{r^2},
\label{eqn:realGSD}
\end{equation}
with two coefficients $B_1$ and $B_2$.
By considering the diagonal and off-diagonal parts of 
eq.~(\ref{eqn:realGSD}) separately, we have 
\begin{align}
2B_1+B_2
&=
\frac{1}{2\pi\eta} \int_0^\infty {\rm d}k \,
\frac{J_0(kr)}{k + \nu} 
\\
&= \frac{1}{4\eta} 
\left[ {\bf H}_0(\nu r) - Y_0(\nu r) \right], 
\label{eqn:saf2AB}
\end{align}
and 
\begin{align}
B_1+B_2
&= \frac{1}{2\pi\eta} 
\int_0^\infty {\rm d}k \,
\frac{J_1(kr)}{kr(k + \nu)}
\\
&= \frac{1}{4\pi\eta} 
\int_0^\infty {\rm d}k \,
\frac{J_0(kr)+J_2(kr)}{k + \nu}
\\
&= \frac{1}{4\eta} 
\left[
-\frac{2}{\pi\nu^2r^2}
+\frac{{\bf H}_1(\nu r)}{\nu r} -\frac{Y_1(\nu r)}{\nu r}
\right].
\label{eqn:safAB}
\end{align}
See ref.~\cite{gradshteyn} for the integral of 
eq.~(\ref{eqn:saf2AB}).
In evaluating the integral of eq.~(\ref{eqn:safAB}), we have made 
use of the following relations;
\begin{equation}
\frac{J_1(z)}{z} = J_0(z) + 
\frac{{\rm d}^2}{{\rm d} z^2} J_0(z),
\end{equation}
\begin{equation}
\int_0^\infty {\rm d}z \,
J_0(z) \frac{{\rm d}^2}{{\rm d}z^2} \frac{1}{z+a}
= \frac{{\rm d}^2}{{\rm d}a^2}\int_0^\infty {\rm d}z \,
\frac{J_0(z)}{z+a}.
\end{equation}
Solving eqs.~(\ref{eqn:saf2AB}) and~(\ref{eqn:safAB}) for 
$B_1$ and $B_2$, we get
\begin{align}
B_1 = & \frac{1}{4\eta} \left[ {\bf H}_0(\nu r) - Y_0(\nu r)
+\frac{2}{\pi\nu^2r^2}
\right.
\nonumber \\
& \left.
-\frac{{\bf H}_1(\nu r)}{\nu r} 
+\frac{Y_1(\nu r)}{\nu r}\right], 
\\
B_2 = & \frac{1}{4\eta} \left[
-\frac{4}{\pi\nu^2r^2}
+\frac{2{\bf H}_1(\nu r)}{\nu r}
\right.
\nonumber \\ 
& \left.
-\frac{2Y_1(\nu r)}{\nu r}
-{\bf H}_0(\nu r) + Y_0(\nu r)
\right].
\end{align}

The preaveraging of eq.~(\ref{eqn:realGSD}) yields 
\begin{equation}
g^{\rm SD}(n-m) = \frac{1}{2} \langle 2 B_1 + B_2 \rangle. 
\end{equation}
Hence we obtain eq.~(\ref{eq:exvSD}) using eq.~(\ref{eqn:saf2AB}).

%%%%%%%%%%%%%%%%%%%%%%%%%%%%%%%%%%%%%%%%%%%%%%%%%%%%
\subsection*{Confined membrane case}

Now the mobility tensor in the Fourier space is 
eq.~(\ref{eqn:esoseen}).
Similar to the free membrane case, its real space expression can be written as
\begin{equation}
G^{\rm ES}_{\alpha\beta}({\bf r}) = 
C_1 \delta_{\alpha\beta} + C_2 \frac{r_\alpha r_\beta}{r^2},
\label{eqn:realGES}
\end{equation}
with two coefficients $C_1$ and $C_2$.
Then it follows that  
\begin{align}
2C_1+C_2 &=
\frac{1}{2\pi\eta} 
\int_0^\infty {\rm d}k \,
\frac{k J_0(kr)}{k^2 + \kappa^2}
\\
&= \frac{1}{2\pi\eta} K_0(\kappa r),
\label{eqn:es2AB}
\end{align}
and 
\begin{align}
C_1+C_2 
&= \frac{1}{2\pi\eta}
\int_0^\infty {\rm d}k \,
\frac{J_1(kr)}{r(k^2 + \kappa^2)}
\\
&= \frac{1}{2\pi\eta} 
\left[ \frac{1}{\kappa^2r^2} - \frac{K_1(\kappa r)}{\kappa r} \right].
\label{eqn:esAB}
\end{align}
Solving these equations, we obtain  
\begin{align}
C_1 & = 
\frac{1}{2\pi\eta} \left[ K_0(\kappa r) 
+ \frac{K_1(\kappa r)}{\kappa r} 
- \frac{1}{\kappa^2 r^2} \right],\\
C_2 &= \frac{1}{2\pi\eta} \left[ -K_0(\kappa r) 
- \frac{2K_1(\kappa r)}{\kappa r} 
+ \frac{2}{\kappa^2 r^2} \right].
\end{align}

%%%%%%%%%%%%%%%%%%%%%%%%%%%%%%%%%%%%%%%%%%%%%%%%%%%%
%\bibliographystyle{epj}
%\bibliography{../../../writeups/refs}

%%%%%%%%%%%%%%%%%%%%%%%%%%%%%%%%%%%%%%%%%%%%%%%%%%%%%
%
\end{document}